\documentclass[showpacs,aps,pre,amsmath,superscriptaddress,longbibliography,twocolumn]{revtex4-1}

\usepackage{mathrsfs}
\usepackage{amssymb}
\usepackage{amstext}
\usepackage{mathtools}
\usepackage[pdftex]{color}
\usepackage[english]{babel}
\usepackage{graphicx}\usepackage{bm}\usepackage{float}
\usepackage[normalem]{ulem}
\usepackage{algorithm}
\usepackage{algpseudocode}
\usepackage[breaklinks]{hyperref}
\hypersetup{colorlinks=true, linkcolor=blue, citecolor=blue, filecolor=blue, urlcolor=blue}

\makeatletter

\newcommand{\Rmnum}[1]{\expandafter\@slowromancap\romannumeral #1@}
\makeatother

\begin{document}

\title{Tensor network method for reversible classical computation}

\author{Zhi-Cheng Yang}

\affiliation{Physics Department, Boston University, Boston,
  Massachusetts 02215, USA}

\author{Stefanos Kourtis}

\affiliation{Physics Department, Boston University, Boston,
  Massachusetts 02215, USA}

\author{Claudio Chamon}

\affiliation{Physics Department, Boston University, Boston,
  Massachusetts 02215, USA}

\author{Eduardo R. Mucciolo}

\affiliation{Department of Physics, University of Central Florida,
  Orlando, Florida 32816, USA}

\author{Andrei E. Ruckenstein}

\affiliation{Physics Department, Boston University, Boston,
  Massachusetts 02215, USA}

\date{\today}

\begin{abstract}
  We develop a tensor network technique that can solve universal
  reversible classical computational problems, formulated as
  vertex models on a square lattice [Nat. Commun. {\bf8}, 15303
    (2017)]. By encoding the truth table of each vertex constraint in
  a tensor, the total number of solutions compatible with partial
  inputs/outputs at the boundary can be represented as the full
  contraction of a tensor network. We introduce an iterative
  compression-decimation (ICD) scheme that performs this contraction
  efficiently. The ICD algorithm first propagates local constraints to
  longer ranges via repeated contraction-decomposition sweeps over all
  lattice bonds, thus achieving compression on a given length
  scale. It then decimates the lattice via coarse-graining tensor
  contractions. Repeated iterations of these two steps gradually
  collapse the tensor network and ultimately yield the exact tensor
  trace for large systems, without the need for manual control of
  tensor dimensions. Our protocol allows us to obtain the exact number
  of solutions for computations where a naive enumeration would take
  astronomically long times.
\end{abstract}

\maketitle

\section{Introduction}

Physics-inspired approaches have led to efficient algorithms for
tackling typical instances of hard computational problems, shedding
new light on our understanding of the complexity of such
problems~\cite{SP, mezard}. The conceptual framework of these
approaches is based on the realization that the solutions of certain
computational problems are encoded in ground states of appropriate
statistical mechanics models. However, the existence of either a
thermodynamic phase transition into a glassy phase or a first-order
quantum phase transition represent obstructions to reaching the ground
state, often even for easy problems~\cite{ricci, thomas, young, hen,
  farhi}. Recently, Ref.~\cite{chamon} introduced a new class of
problems by mapping a generic reversible classical computation onto a
two-dimensional vertex model with appropriate boundary conditions. The
statistical mechanics model resulting from this mapping displays no
bulk thermodynamic phase transitions and the bulk thermodynamics is
independent of the classical computation represented by the
model. Taken together, these features remove an obvious obstacle to
reaching the ground state of a large class of computational problems
and imply that the time-to-solution and the complexity of the problem
are determined by the dynamics of the relaxation of the corresponding
system to its ground state. However, when thermal annealing is
employed, the resulting dynamics is found to be extremely
slow, and even easy computational problems cannot be efficiently
solved. Since \textit{any} classical computation implemented as a
reversible circuit can be formulated in this fashion, finding an
algorithm that can solve the resulting vertex models efficiently
would have far-reaching repercussions.

In this paper, we introduce a tensor network approach that can treat
vertex models encoding computational problems. Tensor networks are a
powerful tool in the study of classical and quantum many-body systems
in two and higher spatial dimensions, and are also used as compressed
representations of large-scale structured data in ``big-data''
analytics~\cite{cichocki1,vervliet,cichocki2}. Here we are interested
in taking the trace of tensor
networks~\cite{biamonte1,biamonte2,biamonte3}, to count the number of
solutions of a computational problem. As opposed to thermal annealing,
which serially visits individual configurations, tensor network
schemes sum over all configurations simultaneously. As a result,
tensor-based approaches lead to a form of virtual parallelization
\cite{chamon2}, which, under certain circumstances, speeds up the
computation of the trace. Most of the physics-driven applications have
focused on tensor network renormalization group (TNRG) algorithms that
coarse grain the network while optimally removing short-range
entanglement~\cite{verstraete,
  levin,gu,Jiang2008,wen2,vidal1,xiang1,vidal2,xiang2,wen,verstraete2,
  kagome,evenbly,evenbly2}. There are, however, two aspects of our
physics-motivated work that are qualitatively different from that of
TNRG approaches. First, vertex models of computational gates are
intrinsically not translationally invariant. Second, the trace over
the tensor network, which counts the number of solutions of the
computational circuit (the analogue of the zero temperature partition
function of statistical mechanics models) must be computed exactly,
within machine precision. (Approximations of the tensors lead to
approximate counting, which in certain problems is no easier than
exact counting~\cite{Feige1991}.) Both features are naturally treated
by the methods proposed in this paper.

In our tensor network approach, the truth table of each vertex
constraint corresponding to a computational gate is encoded in a
tensor, such that the local compatibility between neighboring bits (or
spins) is automatically guaranteed upon contracting the shared bond
between two tensors. Summing over all possible unfixed boundary vertex
states and contracting the entire tensor network give the partition
function, which counts the total number of solutions compatible with
the boundary conditions, a problem belonging to the class \#P. Finding
\textit{a} solution can then be accomplished by fixing one boundary
vertex at a time, with the total number of trials linear in the number
of input bits~\cite{chamon2}.

Our tensor network method, which we refer to as the iterative
compression-decimation (ICD) algorithm, can be regarded as a set of
local moves defining a novel dynamical path to the ground state of
generalized vertex models on a square lattice. These moves can be
shown to decrease or leave unchanged the bond dimensions of the
tensors involved, thus achieving optimal compression (i.e., minimal
bond dimension) of the tensor network on a lattice of fixed size. The
algorithm's first step is to propagate local vertex constraints across
the system via repeated contraction-decomposition sweeps over all
lattice bonds. These back and forth sweeps are the higher dimensional
tensor-network analog of those employed in the one-dimensional
finite-system density matrix renormalization group (DMRG)
method~\cite{DMRG}. For problems with non-trivial boundary conditions,
such as those encountered in computation, these sweeps also propagate
the boundary constraints into the bulk, thus progressively building
the connection between opposite (i.e., input/output) boundaries. In
the next step, the algorithm decreases the size of the lattice by
coarse-graining the tensor network via suitable contractions. Repeated
iterations of these two steps allow us to reach larger and larger
system sizes while keeping the tensor dimensions under control, such
that ultimately the full tensor trace can be taken.

The computational cost of ICD hinges upon the maximum bond dimension
of the tensors during the coarse-graining procedure. We identify the
hardness of a given counting problem by studying the scaling of the
maximum bond dimension as a function of the system size, the
concentration of nontrivial constraints imposed by TOFFOLI gates, and
the ratio of unfixed boundary vertices. We further present both the {\it average} and {\it typical} maximum bond dimension distributions over random instances of computations.
While we cannot distinguish
between polynomial and exponential scaling for the hardest regime of
high TOFFOLI concentration, there exist certain regimes of the problem
where the bond dimension grows relatively slowly with system
size. Therefore, within this regime, we are able to count the exact
number of solutions within a large search space that is intractable
via direct enumerations.

The rest of the paper is organized as follows. We first briefly
introduce the tensor network representation of generic vertex models
on a square lattice in
Sec.~\ref{sec:vertex}. Section~\ref{sec:algorithm} describes the ICD
algorithm for coarse-graining and efficiently contracting the tensor
network. In Section~\ref{sec:toffoli} we apply the ICD method to
reversible classical computational problems as encoded in the vertex
model of computation introduced in Ref.~\cite{chamon}, and discuss a relation between
the number of solutions of the computational problem and the maximum
bond dimension of the tensor network from an entanglement
perspective. Section~\ref{sec:random} presents the implementation of the ICD and the accompanying numerical scaling
results for random computational networks defined by the concentration of Toffoli gates placed randomly at vertices of a tilted planar square lattice.
Finally, we close with Sec.~\ref{sec:outlook}, where we
outline future applications of our ICD algorithm to both computational
and physics problems. 

\section{Tensor network for vertex models}
\label{sec:vertex}

\begin{figure}[hbt]
\centering
\includegraphics[width=.45\textwidth]{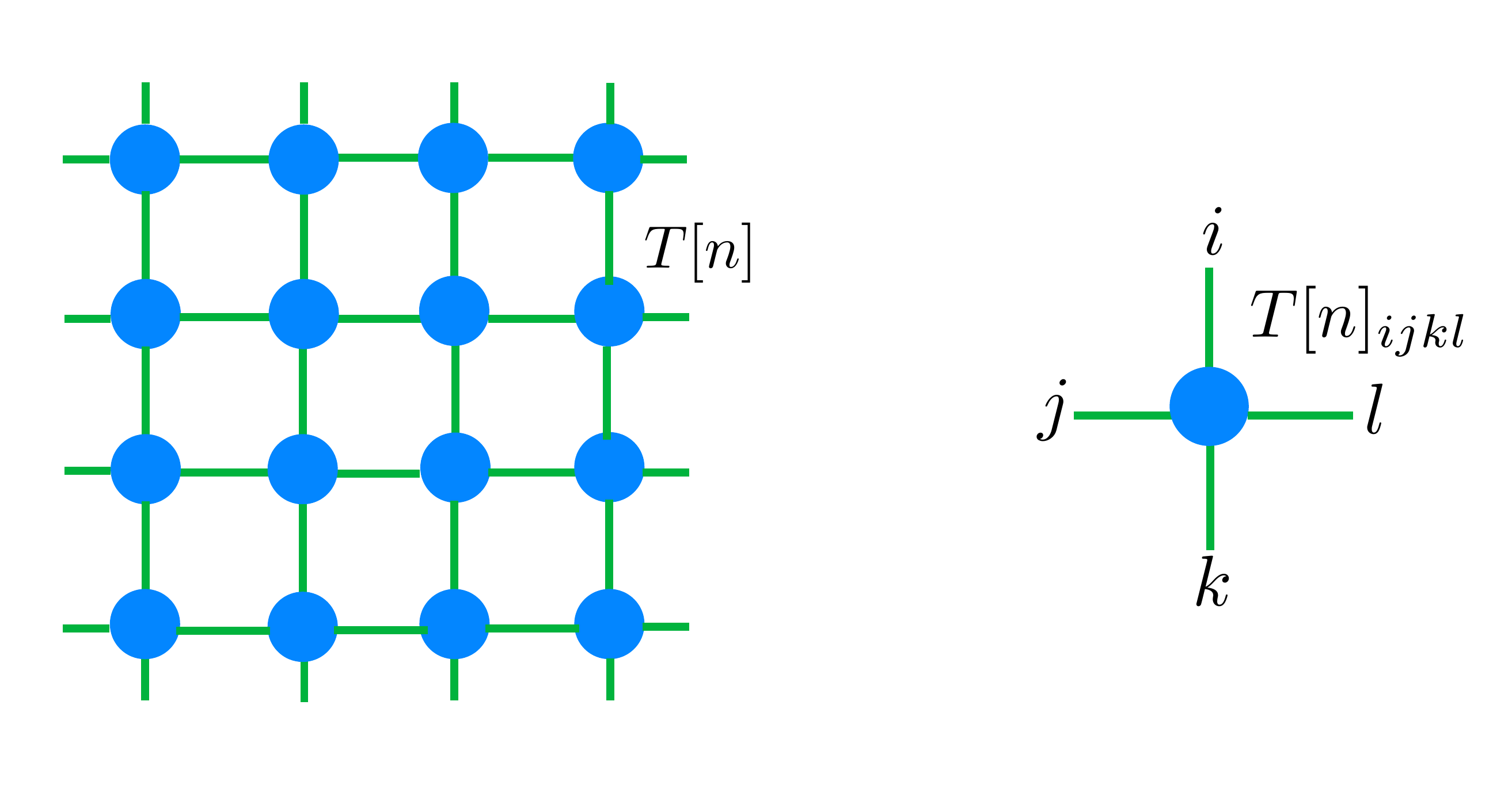}
\caption{(Color online) Vertex model on a square lattice. A local
  tensor $T[n]_{ijkl}$ is defined on each lattice site.}
\label{fig:tensor_net} 
\end{figure}

We start by introducing the tensor network representation for a
generic vertex model. In our formulation, discrete degrees of freedom
reside on the \textit{edges} of a regular lattice and they are coupled
locally to their neighboring degrees of freedom. Couplings between
degrees of freedom are denoted by \textit{vertices}. The couplings at
each vertex $n = 1, \ldots, N_{\mathrm{sites}}$, where
$N_{\mathrm{sites}}$ denotes the total number of vertices, are encoded
into a tensor $T[n]$ whose rank will depend on the connectivity of the
lattice. Fixing the state at all edges incident to a vertex collapses
the corresponding tensor to a scalar. For concreteness, let us
consider the square lattice as an example, as shown in
Fig.~\ref{fig:tensor_net}; generalizations to other types of lattices
are straightforward. Each tensor $T[n]$ is therefore a rank-4 tensor
$T[n]_{ijkl}$, where $i, j, k, l$ denote bond indices.

The tensor representation is quite general. If, for example, one
associates a Boltzmann weight with each combination of bond index
values, one can encode statistical mechanics problems into the tensor
network~\cite{levin, gu, xiang1, vidal2, xiang2, wen}. Alternatively,
by assigning boolean 1s to ``compatible'' combinations of bond index
values and boolean 0s to ``incompatible'' ones, such that the tensor
represent a vertex constraint or a truth table, one can either study
statistical mechanical vertex models at zero temperature,
or implement computational circuits with the tensor network
(Sec.~\ref{sec:toffoli}). Finally, one could even embed the weights of
a discretized path integral for a 1+1D quantum problem in a
two-dimensional network. For finite systems with boundaries, the
boundary tensors will have a different rank from the bulk tensors.

We define the tensor trace of the network as
\begin{equation}
Z = {\rm tTr} \prod_n T[n]_{ijkl},
\label{eq:partition}
\end{equation}
where $n$ runs over all lattice sites and tTr denotes full
contractions of all bond indices. This trace may correspond to the
partition function for the 2D classical system, or the number of
possible solutions of a computation, or the imaginary-time path
integral for a 1D quantum system, etc. In general, a brute-force
evaluation of the full tensor trace multiplies the dimensions of the
tensors, thereby requiring a number of operations exponential in
system sizes. It is therefore expedient for any strategy of evaluating
the trace to keep the dimensions of tensors under control at
intermediate steps, so that the tensor trace can be ultimately
taken. Ideally, one would like a protocol that uses all available
information --- such as boundary conditions, compatibility
constraints, energy costs or Boltzmann weights, depending on the
particular problem at hand --- to compress the tensor network as much
as possible, while maintaining all the essential information
therein. In Sec. \ref{sec:algorithm}, we propose an efficient
iterative scheme that achieves this goal.  In particular, as we detail
in Sec.~\ref{sec:toffoli}, our algorithm provides a simple way to deal with
finite systems without translational invariance, and subject to
various types of boundary conditions.
 
\section{compression-decimation algorithm}
\label{sec:algorithm}

In this section, we describe the compression-decimation algorithm that
facilitates the exact contraction of tensor networks. The algorithm
consists of two steps. First, we perform \textit{sweeps} on the
lattice via a singular value decomposition (SVD) of pairs of tensors
in order to eliminate short-range entanglement and propagate
information from the boundary to the bulk, hence removing the
redundancies in the bond dimensions. Due to its nature, we call this
step \textit{compression}. Next, we contract pairs of rows and columns
of the lattice such that the system size is reduced. This step is
referred to as \textit{decimation}. The two steps are then repeated
until the size and bond dimensions of the tensor network become small
enough to allow an exact full contraction of the network.

Locally, the sweeps remove redundancies due to either short-range
entanglement or incompatibility in the local tensors, and compress the
information into tensors with smaller bond dimensions. Globally, the
sweeps propagate information about the boundary conditions to the
bulk, thus imposing global constraints on the local bulk
tensors. Moreover, since the sweeping is performed back and forth
across the entire lattice, it does not differentiate between whether
or not translational invariance is present. Therefore, our scheme may
be thought of as a higher dimensional analog of the finite-system DMRG
algorithm that applies to generic vertex models on finite lattices.

\subsection{Compression}\label{sec:algorithm-sweeps}

In this step, we visit sequentially each bond in the lattice and
contract the corresponding indices of the two tensors sharing this
bond. We then perform an SVD on the contracted bond and truncate the
singular value spectrum keeping only those greater than a certain
threshold $\delta$. After that, the tensors are reconstructed with a
smaller bond dimension. We define each forward plus backward traversal
of all the bonds in the network as one \textit{sweep}. The specific
choice of the threshold $\delta$ depends on the desired precision, as
well as the problem we are dealing with. For example, in formulating
TNRG algorithms, $\delta$ can be chosen to be some small but finite
number. On the other hand, for computational problems such as
counting, $\delta$ is chosen to be zero within machine precision. 

\begin{figure}[hbt]
\centering
\includegraphics[width=.47\textwidth]{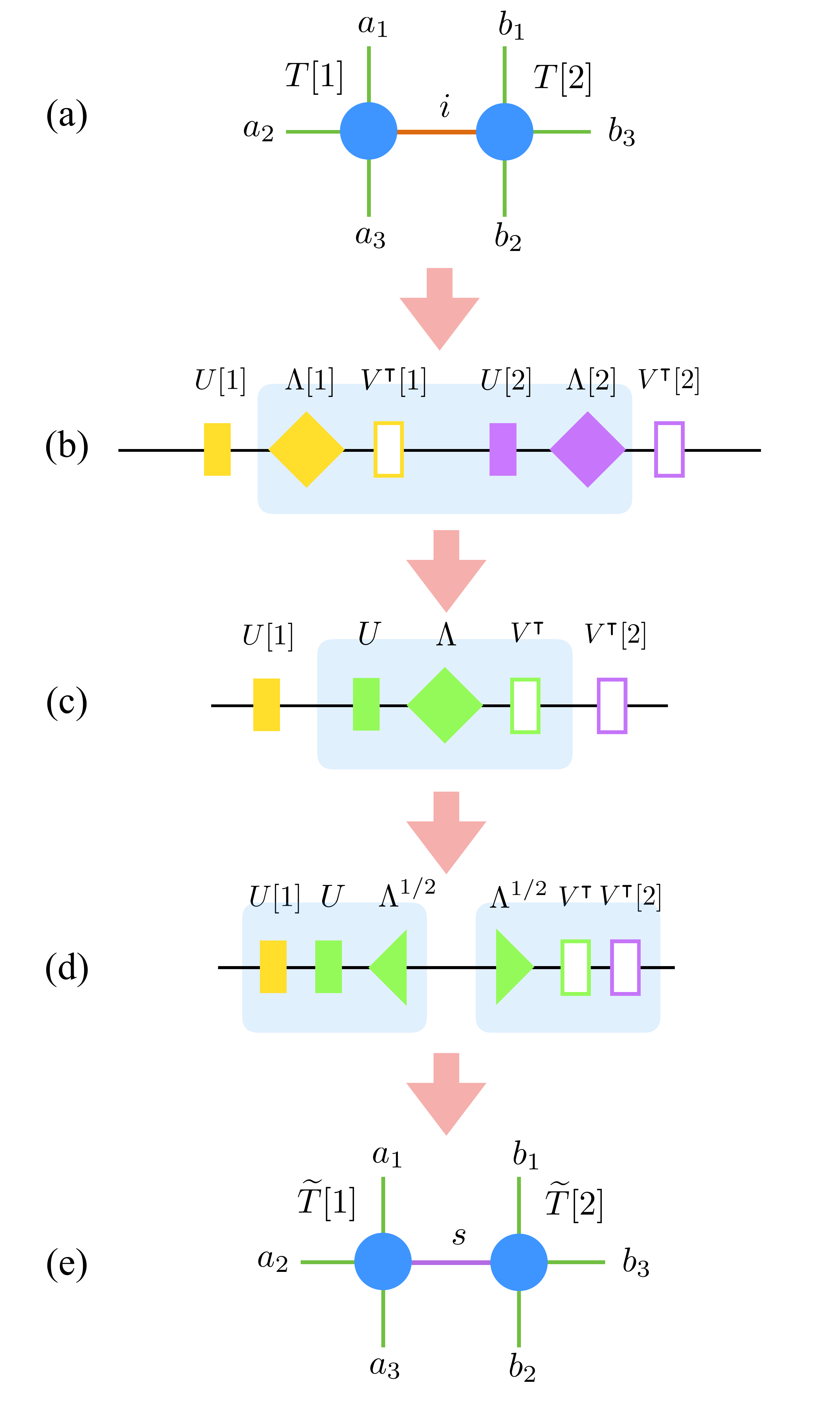}
\caption{(Color online) The contraction-decomposition step in the
  sweeping. (a) Two tensors $T[1]$ and $T[2]$ sharing a bond $i$. (b)
  Perform SVDs on individual tensors respectively. (c) Perform an SVD
  on the shaded part. (d) Split the resultant matrices into two
  pieces. (e) Construct new tensors $\widetilde{T}[1]$ and
  $\widetilde{T}[2]$.}
\label{fig:sweep} 
\end{figure}

Let us take two tensors with the shared bond labeled by $i$,
$T[1]_{a_1 a_2 a_3 i}$ and $T[2]_{b_1 i b_2 b_3}$, as shown in
Fig.~\ref{fig:sweep}a, where we denote the dimension of bond $i$ as
$d_i$. We would like to reduce $d_i$ via a SVD. In
principle, this can be achieved by directly contracting $T[1]$ and
$T[2]$ along dimension $i$ into a matrix $M_{A,B}=T[1]_{A, i}
T[2]_{i,B}$, where we have grouped the other three indices of each
tensor into superindices $A \equiv (a_1a_2a_3)$ and $B \equiv
(b_1b_2b_3)$, and then performing an SVD. However, to avoid decomposing
the matrix $M_{A,B}$ with potentially large bond dimensions, we first
do an SVD on each individual tensor (Fig.~\ref{fig:sweep}b):
\begin{subequations}
\begin{align}
T[1]_{A,i} &=& U[1]_{A,r}\, \Lambda[1]_{r}\, V[1]^\intercal_{r,i}  \,,\\
T[2]_{i,B} &=& U[2]_{i,r'}\, \Lambda[2]_{r'}\, V[2]^\intercal_{r',B}.
\end{align}
Notice that the contraction of $T[1]$ and $T[2]$ can then be written
as
\begin{equation}
T[1] T[2] = U[1]_{A,r} \bigg[ \Lambda [1]_{r} V[1]^\intercal_{r,i}
  U[2]_{i,r'} \Lambda[2]_{r'} \bigg] V[2]^\intercal_{r',B}.
\label{eqn:contraction}
\end{equation}
This implies that we can instead perform an SVD on the part shown
within brackets in Eq.~(\ref{eqn:contraction}): $\widetilde{M}_{r,r'}
= \Lambda[1]_{r} V[1]^\intercal_{r,i} U[2]_{i,r'} \Lambda[2]_{r'}$,
which has much smaller dimensions since $d_r \leq {\rm min}(d_A, d_i),
d_{r'} \leq {\rm min}(d_B, d_i)$. Now we perform an SVD on the matrix
$\widetilde{M}_{r,r'}$ to obtain (Fig.~\ref{fig:sweep}c)
\begin{equation}
\widetilde{M}_{r,r'} = U_{r,s} \Lambda_{s} V^\intercal_{s,r'}.
\end{equation}
(At each SVD step described above, we discard singular values that are
smaller than $\delta$.) Therefore, after the above steps, the bond
dimension $d_s\leq{\rm min}(d_r, d_{r'})\leq {\rm min}(d_i, d_A,
d_B)$. Finally, we construct new tensors as 
\begin{align}
\widetilde{T}[1]_{a_1 a_2 a_3 s} \equiv &{\ } \widetilde{U}_{(a_1 a_2
  a_3), s}\, (\Lambda_{s})^{1/2} \,,\\ \widetilde{T}[2]_{s b_1 b_2 b_3}
\equiv &{\ } (\Lambda_{s})^{1/2}\, \widetilde{V}^\intercal_{s, (b_1 b_2
  b_3)} \,,
\end{align}\label{eq:sweeps}\end{subequations}
where the dimension of the shared bond is reduced
(Fig.~\ref{fig:sweep}d,e). Starting from one boundary, we visit
sequentially each bond $i \in 1, \ldots, N_{\mathrm{bonds}}$, where
$N_{\mathrm{bonds}}$ is the total number of bonds in the lattice, and
perform the steps outlined above, until we reach the opposite
boundary. Then we repeat the procedure in the opposite direction,
until we reach the original boundary. The sweeping can be repeated
$N_{\mathrm{sweeps}}$ times, or until convergence of all bond
dimensions.

\subsection{Decimation}
\label{sec:coarse-graining}

\begin{figure}[t]
\centering
\includegraphics[width=.47\textwidth]{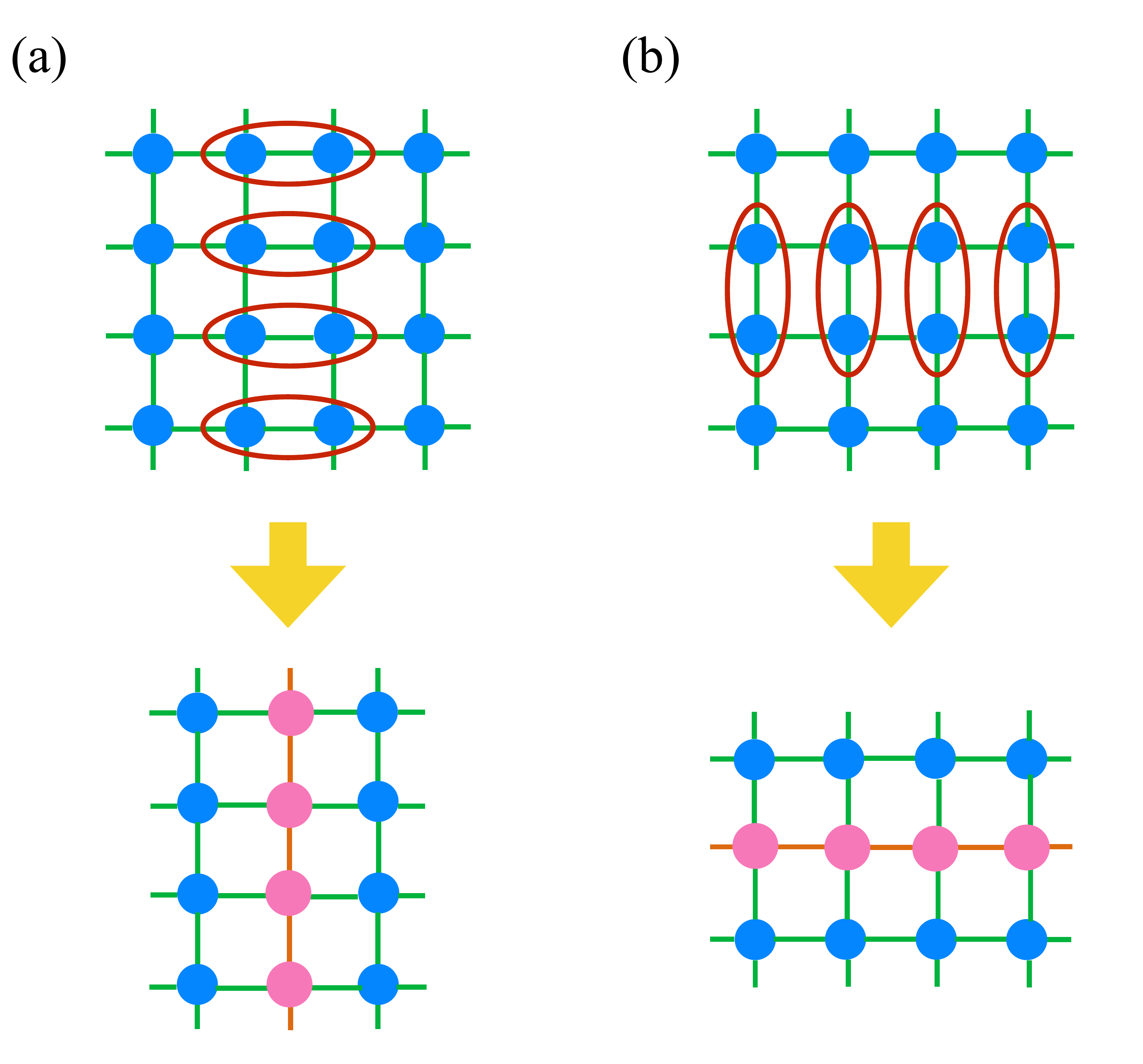}
\caption{(Color online) (a) A column contraction involving pairs of
  tensors along the $x$ direction. (b) A row contraction involving
  pairs of tensors along the $y$ direction. The new tensors resulting
  from the contractions are denoted by pink dots, and the new bonds
  are denoted by orange lines.}
\label{fig:coarse_grain} 
\end{figure}

The second step of the algorithm is to contract pairs of rows and
columns of the tensor network, so as to yield a lattice with a smaller
number of sites~\cite{xiang1, xiang2}. As we show in
Fig.~\ref{fig:coarse_grain}, this step consists of a column
contraction (Fig.~\ref{fig:coarse_grain}a), followed by a row
contraction (Fig.~\ref{fig:coarse_grain}b). In the column contraction,
we contract pairs of tensors along the $x$ direction, and obtain a new
tensor (see also Fig.~\ref{fig:sweep}a):
\begin{equation}
  \mathcal{T}_{(a_1b_1)a_2(a_3b_2)b_3} = \sum_i T[1]_{a_1 a_2 a_3 i}\,
  T[2]_{b_1 i b_2 b_3}.
  \label{eq:column-contraction}
\end{equation}
We then perform a row contraction similarly, during which pairs of
tensors are contracted along the $y$ direction.

During this step, the dimensions of the bonds perpendicular to the
current direction of contractions are multiplied and hence will
inevitably grow. Therefore, after all columns/rows are contracted, we
sweep back and forth again to reduce the bond dimensions. A simplified
version of the compression-decimation scheme is presented as
pseudocode in Algorithm~\ref{alg:cd}.

\begin{algorithm}[H]
\caption{\label{alg:cd}Iterative Compression-Decimation}
\textbf{Input:} tensor network on a square lattice $\lbrace T[n] \, |
\, n \in 1,\dots,N_{\mathrm{sites}} \rbrace$; $N_{\mathrm{sweeps}}\geq
1$; $\delta\geq 0$ (SVD truncation parameter).\\ \textbf{Output:} $Z$,
as defined in Eq.~\eqref{eq:partition}
  \begin{algorithmic}[1]
  \Repeat
    \For{$i=1,\ldots,N_{\mathrm{sweeps}}$}{\ (compression)}
    \For{$b=1,\ldots,N_{\mathrm{bonds}}$}{\ (forward sweep)}
    \State Contract, SVD, and update tensors as in Eq.~\ref{eq:sweeps}
    \EndFor
    \For{$b=N_{\mathrm{bonds}},\ldots,1$}{\ (backward sweep)}
    \State Carry out backward sweep similarly
    \EndFor
    \EndFor
    \State Perform column contractions by Eq.\eqref{eq:column-contraction} (decimation)
    \State Perform row contractions similarly (decimation)
  \Until{network is decimated to single site}
  \State Carry out tensor trace Eq.~\eqref{eq:partition}
  \end{algorithmic}
\end{algorithm}

A few remarks are in order. First, the lattice structure lends us more
flexibility with the coarse-graining step since one does not have to
contract every pair of rows and columns. For example, in cases of
systems without translational invariance, representing either
disordered statistical mechanics models or models encoding
computational circuits, the bond dimensions are in general not
distributed uniformly across the entire lattice. One could then
perform the contractions selectively on rows and columns containing
mostly tensors with small bond dimensions while leaving the rest for
the next coarse-graining step. In practice, one could set an
appropriate threshold in the algorithm depending on the specific
problems.  Second, the procedure described here is closely related to
the TNRG algorithms where the key is to optimally remove short-range
entanglement at each RG step. For example, Ref.~\cite{wen} proposes a
loop optimization approach for TNRG. An important step in that method
is to filter out short-range entanglement within a plaquette via a QR
decomposition, which we believe should be equivalent to our SVD-based
sweeping. Moreover, as shall be shown in Sec.~\ref{sec:results}, the
sweeps take into account the local environment around each tensor.
The loop structure of short-range entanglement is eliminated (at least
partially) when we visit each bond around the loop and sweep across
the whole system. Whether or not more elaborate schemes~\cite{wen2,
  vidal1,xiang1,vidal2,xiang2,wen,verstraete2,kagome,evenbly,evenbly2,Evenbly2017}
for taking into account the tensor environment can improve the
performance of the sweeps in the ICD scheme will not concern us in
this work: we will see that even the simple sweep protocol described
above is sufficient for the solution of complex generic computational
problems.  Third, our procedure is more apt for systems without
translational invariance, e.g., spin glasses. Finally, the
computational cost scales as $O(\chi^5)$ for the SVD steps, and $O(\chi^7)$ for the
tensor contraction steps, where $\chi$ is the maximum bond dimension of the tensors. 
Hence the computational cost of our compression-decimation algorithm scales as
$O(\chi^7)$.

\section{The general TOFFOLI-based vertex model}
\label{sec:toffoli}

In this section, we provide an example of a hard computational problem
where our scheme can be applied to find solutions in cases that are
otherwise intractable. The models we study here follow from the vertex model
representation of reversible classical computations introduced in
Ref.~\cite{chamon}. We remark that this general vertex model can address 
generic satisfiability problems, a statement that follows from a series of results already documented in the literature:
\begin{enumerate}
\item{The circuit satisfiability (CSAT) problem is NP-complete~\cite{cook, levin2};}
\item{The CSAT problem can be formulated in terms of reversible circuits~\cite{nielsen};}
\item{Any reversible circuit can be constructed using only TOFFOLI gates~\cite{nielsen};}
\item{Any reversible circuit constructed out of TOFFOLI gates can be mapped onto our vertex model
representation, with the addition of an appropriate number of identity and swap gates~\cite{chamon}.}
\end{enumerate}
Hence, our vertex model can encode other satisfiability problems such as 3-SAT, which can
be mapped into CSAT. (Indeed, it is possible to program 3-SAT with $n$ variables and
$m$ clauses into a vertex model using a lattice of size $n \times 2m$~\cite{prep}.)

The vertex model is defined on a square lattice of finite
size with periodic boundary conditions in the transverse direction,
thus placing the model on a cylinder. Depending on the specific computation,
different types of boundary conditions are imposed in the longitudinal
direction. In addition, this model does not display translational
invariance since different gates of the computational circuit are implemented by different vertices. 
This model can encode general computational
problems, including any of the hard instances, and serves as an
excellent candidate to benchmark the performance of our scheme. 

\begin{figure}[hbt]
\centering
\includegraphics[width=.5\textwidth]{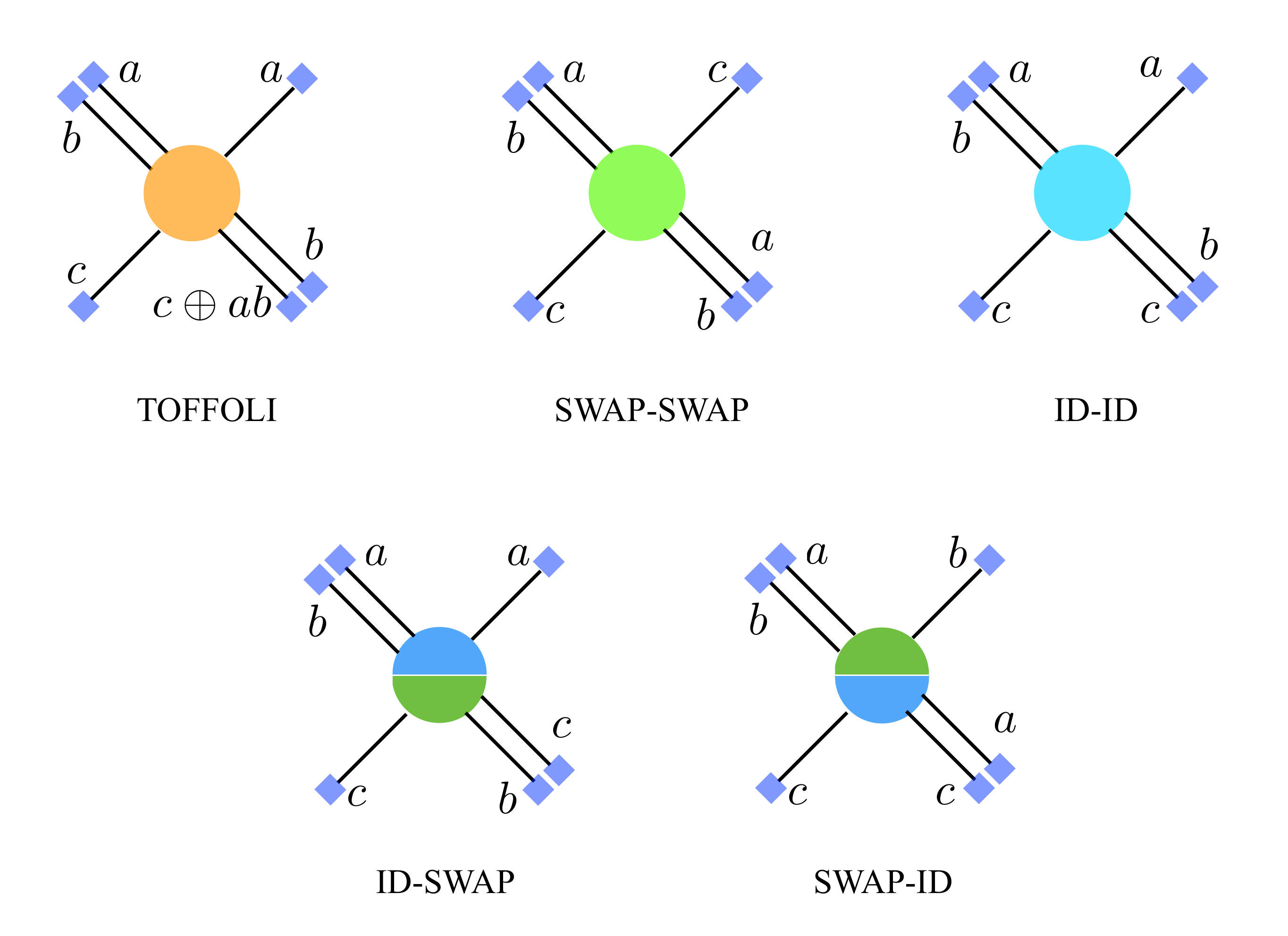}
\caption{(Color online) Five types of vertices used for the vertex
  model representation of reversible classical computations. The input
  and output bits are denoted by blue squares on the links associated
  with a given vertex. }
\label{fig:vertex_gate} 
\end{figure}

We start by giving a self-contained review of the general vertex model
encoding reversible classical computations introduced in Ref.~\cite{chamon} and construct its tensor
network representation. This is based on the fact that any Boolean function can be implemented using a reversible circuit
constructed out of TOFFOLI gates, which are reversible three-bit logic
gates taking the inputs $(a, b, c)$ to $(a, b, ab\oplus c)$. To
facilitate the coupling of far-away bits while maintaining the
locality of TOFFOLI gates, we use two-bit SWAP gates to swap
neighboring bits, $(a, b) \rightarrow (b, a)$, until pairs of distant
bits are adjacent to one another. Bits that do not need to be moved
are simply copied forward using two-bit Identity (ID) gates. To obtain
a plane-covering tiling and thus a square-lattice representation of
the circuit, we combine the SWAP and ID gates into the three-bit gates: ID-ID, ID-SWAP, SWAP-ID, SWAP-SWAP, and represent each of them as
well as the TOFFOLI gate as a vertex with three inputs and three
outputs. The five types of vertices are shown in
Fig.~\ref{fig:vertex_gate}, with the input and output bits explicitly
drawn on the links.

Alternatively, one can think of bits as spin 1/2 particles located on
the bonds between vertices, whereas each vertex imposes local
constraints between ``input'' and ``output'' spins, such that only
$2^3=8$ out of the $2^6=64$ total configurations are allowed. For all
five types of vertices, one can write local one- and two-spin
interaction terms, such that the allowed configurations are given by
the ground-state manifold of the Hamiltonian comprised of all these
terms~\cite{chamon}. The allowed configurations are then separated
from the excited states by a gap set by the energy scale of the
couplings. In the large-couplings limit, interactions can be
equivalently thought of as constraints and one therefore needs only to
consider the subspace where local vertex constraints are always
satisfied.

\begin{figure}[hbt]
\centering
\includegraphics[width=.45\textwidth]{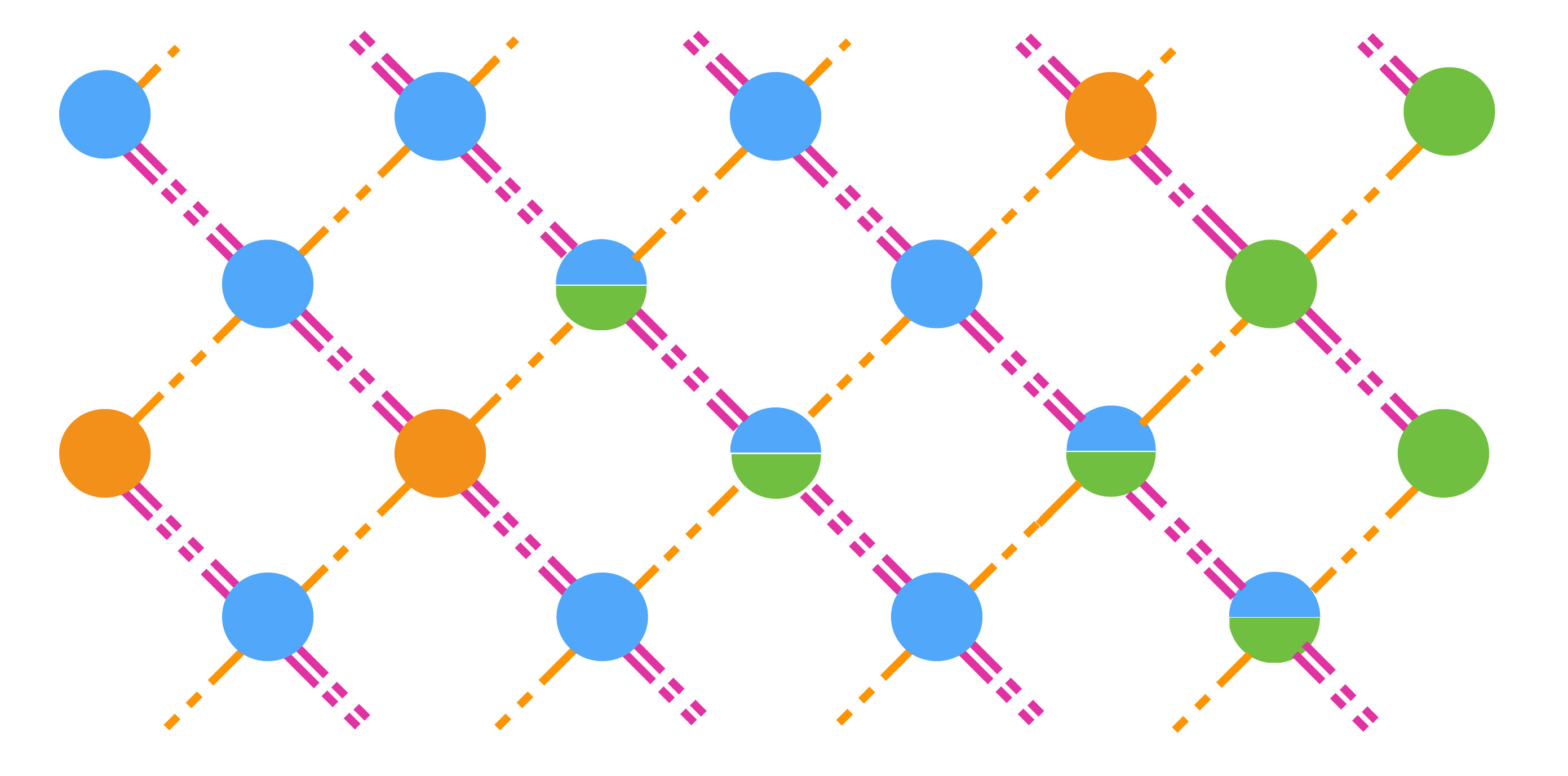}
\caption{(Color online) Vertex model on a tilted square lattice
  encoding a generic classical computation. The left and right
  boundaries stores the input and output states, and periodic boundary
  condition is taken along the transverse direction. }
\label{fig:vertex} 
\end{figure}

Using the five types of vertices introduced above, one can map an
\textit{arbitrary} classical computational circuit onto a vertex model
on a tilted square lattice, as shown in Fig.~\ref{fig:vertex}.

Bits at the left and right boundaries store the input and output
respectively, and the horizontal direction corresponds to the
computational ``time'' direction. The boundary condition along the
transverse direction is chosen to be periodic. Spin degrees of freedom
representing input and output bits associated with each vertex are
placed on the links. This model can be shown to display no
thermodynamic phase transition irrespective of the circuit
realizations via a straightforward transfer matrix
calculation~\cite{chamon}.

When either only the input or only the output boundary bits are fully
determined a priori, the physical system functions as a regular
circuit: the solution can be obtained by passing the boundary state
through the next column of gates, obtaining the output, then passing
this output on to the next column of gates, repeating the procedure
until the other boundary is reached. This mode of solution, which we
shall call \textit{direct computation}, is trivial and its
computational cost scales linearly with the area of the system.

On the other hand, by fixing only a subset of the left and right
boundaries, a class of nontrivial problems can be encoded in the
vertex model. For example, one can cast the integer factorization
problem on a reversible multiplication circuit precisely in this
way~\cite{chamon, vedral}. In these cases, the boundary state cannot
be straightforwardly propagated from the boundaries throughout the
entire bulk, as the input or output of one or more gates is at most
only partly fixed, and therefore direct computation unavoidably
halts. Without any protocol of communication between the two partially
fixed boundaries, one is left with trial-and-error enumeration of all
boundary configurations, whose number grows exponentially with the
number of unfixed bits at the boundaries. Even though it is sometimes
possible to exploit special (nonuniversal) features of specific
subsets of problems in order to devise efficient strategies of
solution (e.g., factorization with sieve algorithms), general schemes
that perform favorably in solving the \textit{typical} instances in
the encompassing class are important, both for highlighting the
underlying universal patterns and as launchpads towards customized
solvers for particular subsets of problems. The algorithm introduced
in this work is of the latter general kind.

\subsection{Tensor network representation}

We shall now construct a tensor network representation of the vertex
model, such that the full contraction of the tensors yields the total
number of solutions satisfying the boundary conditions. In the
statistical mechanics language, this is the partition function of the
vertex model at zero temperature, which essentially counts the ground
state degeneracy.

\begin{figure}[hbt]
\centering
\includegraphics[width=.45\textwidth]{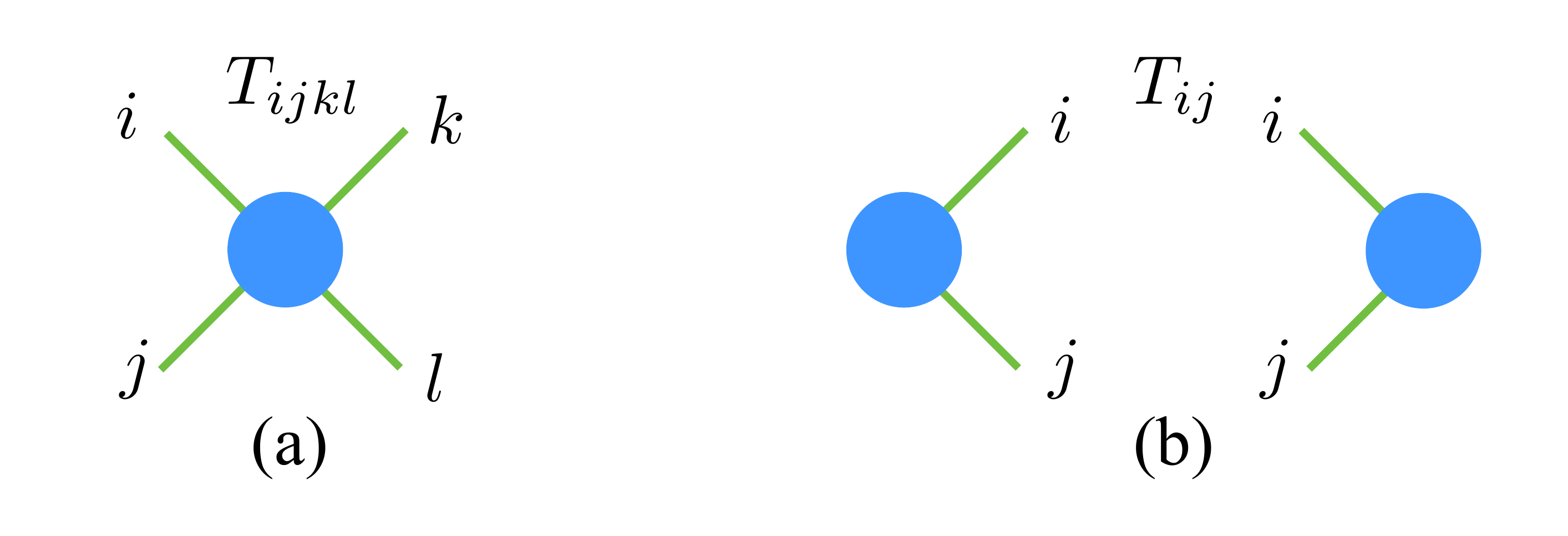}
\caption{(Color online) Definition of (a) bulk and (b) boundary
  tensors.}
\label{fig:tensor} 
\end{figure}

\textit{Bulk tensors.} We define a rank-4 tensor associated with each
vertex in the bulk, $T_{ijkl}$, as shown in
Fig.~\ref{fig:tensor}a. The tensor components are initialized to
satisfy the truth table of the vertex constraint, meaning that
$T_{ijkl}=1$ if $(ij)\rightarrow(kl)$ satisfies the vertex constraint,
and $T_{ijkl}=0$ otherwise. Here the indices should be understood as
integers labeling the spin (bit) states on each bond. Notice that the
indices $i, l$ correspond to double bonds on the lattice while $j, k$
correspond to single bonds. Therefore, the original bond dimensions of
the indices $(i, j, k, l)$ are $(4, 2, 2, 4)$.

For concreteness, let us give an example of encoding the truth table
of the TOFFOLI gate into the tensor $T_{ijkl}$. First, recall that the
gate function of TOFFOLI is $(a, b, c) \rightarrow (a, b, d=c\oplus
ab)$. Comparing Fig.~\ref{fig:tensor}a with
Fig.~\ref{fig:vertex_gate}, we identify on the input side,
$i\equiv(ab) = 2^1b+2^0a, j=c$; on the output side, $k=a, l\equiv(bd)
= 2^1d+2^0b$. In Table~\ref{table}, we explicitly list the truth table
of the TOFFOLI gate and its corresponding non-zero tensor
components. All unspecified tensor components are set to zero. Tensors
encoding the other four types of vertex constraints can be obtained in
a similar fashion.

\begin{table}[hbt]
\centering
\begin{tabular}{c c c | c c c | c}
\hline
\hline
\multicolumn{3}{c|}{input} & \multicolumn{3}{c|}{output} & tensor component \\
\hline
 \ $a$ & $b$ & $c$ & \ $a$ & $b$ & $d$ & $T_{ijkl}\equiv T_{(ab)ca(bd)}$ \\
\hline
\hline
\ 0 & 0 & 0 & \ 0 & 0 & 0 & $T_{0000}=1$ \\
\hline
\ 0 & 0 & 1 & \ 0 & 0 & 1 & $T_{0102}=1$ \\
\hline
\ 0 & 1 & 0 & \ 0 & 1 & 0 & $T_{2001}=1$ \\
\hline
\ 0 & 1 & 1 & \ 0 & 1 & 1 & $T_{2103}=1$ \\
\hline
\ 1 & 0 & 0 & \ 1 & 0 & 0 & $T_{1010}=1$ \\
\hline
\ 1 & 0 & 1 & \ 1 & 0 & 1 & $T_{1112}=1$ \\
\hline
\ 1 & 1 & 0 & \ 1 & 1 & 1 & $T_{3013}=1$ \\
\hline
\ 1 & 1 & 1 & \ 1 & 1 & 0 & $T_{3111}=1$ \\
\hline
\hline
\end{tabular}
\caption{Truth table and the corresponding tensor components for the
  TOFFOLI gate. On the input side, $i\equiv(ab) = 2^1b+2^0a, j=c$; on
  the output side, $k=a, l\equiv(bd) = 2^1d+2^0b$. All unspecified
  components are zero.}
\label{table}
\end{table}

\textit{Boundary tensors.} The vertices at the boundary have only two
bonds. Hence we define a rank-2 tensor $T_{ij}$ at the boundary, where
the indices $i,j$ have the same meaning as the bulk tensors
(Fig.~\ref{fig:tensor}b). Here we draw a distinction between boundary
tensors whose vertex states are fixed and those that are not. For
fixed boundary vertices, $T_{ij}=1$ only for one component
corresponding to the fixed state, whereas for unfixed ones, $T_{ij}=1
\, \forall \, i,j$.

Under the above definitions of local tensors, local compatibility
between spins shared by two vertices is automatically guaranteed when
contracting the corresponding two tensors. Moreover, the unfixed
boundary tensors already encode the information of all possible vertex
states in a compact way, fulfilling a form of classical virtual
parallelization~\cite{chamon2}. Therefore, the full contraction of the
tensor network --- if it can be performed --- will give the total
number of solutions subject to a certain boundary condition.

\subsection{Entanglement and number of solutions}
\label{sec:entanglement}

Before moving on to the concrete application of the algorithm, let us
try to gain some insights into the bond dimensions of local tensors
needed to encode the information of the total number of solutions from
an entanglement point of view~\cite{chamon3}. Let us denote the
collection of free vertex states at the input and output boundaries by
$\{q_{{\rm in}}\}$ and $\{q_{{\rm out}}\}$. We construct a weight
$W(\{q_{{\rm in}}, q_{{\rm out}}\})$ which equals 1 if the state
$\{q_{\rm in}, q_{\rm out}\}$ is a solution, and 0 otherwise. The
partition function is then given by
\begin{equation}
Z = \sum_{\{q_{\rm in}, q_{\rm out}\}} W(\{q_{{\rm in}}, q_{{\rm
    out}}\}),
\end{equation}
which equals the total number of solutions. Now we construct a
\textit{quantum} state as follows:
\begin{widetext}
\begin{eqnarray}
|\psi\rangle & = & \sum_{\{q_{{\rm in}}, q_{{\rm out}}\}} W(\{q_{{\rm
    in}}, q_{{\rm out}}\}) \ |\{q_{{\rm in}}, q_{{\rm out}}\}\rangle
\nonumber \\ &=& \sum_{\{q_{{\rm in}}, q_{{\rm out}}\}} {\rm
  tTr}\left(T[1]^{q_{\rm in 1}}T[2]^{q_{\rm in 2}} \dotsb T[i] T[i+1]
\dotsb T[N]^{q_{{\rm out} L_{\partial}}}\right ) \ |\{q_{{\rm in}}\},
\{q_{{\rm out}}\}\rangle,
\label{eqn:quantum}
\end{eqnarray}
\end{widetext}
where $N$ is the total number of vertices, $L_{\partial}$ is the
number of unfixed vertices on each boundary, and tTr denotes tracing
over all internal indices of the tensors. Let us imagine taking a cut
perpendicular to the periodic direction and divide the system into two
subsystems. The entanglement between subsystem left and right is
determined by the singular value spectrum of the \textit{matrix}
$\mathcal{W}(\{q_{{\rm in}}\}, \{q_{{\rm out}}\})$ reshaped from the
weight $W(\{q_{{\rm in}}, q_{{\rm out}}\})$. $\mathcal{W}$ is a matrix
whose entries are either 1 or 0, and there can be at most one entry in
each row and column that equals 1 due to the reversible nature of the
circuit. Thus the rank of the matrix $\mathcal{W}$ is $Z$, and the
zeroth-order R\'enyi entropy $S^{(0)}={\rm ln}Z$. The entanglement
entropy of the quantum state~(\ref{eqn:quantum}) is hence upper
bounded by $S^{(1)} \leq S^{(0)} = {\rm ln}Z$. Therefore, at least
when there is only a small (nonextensive) number of solutions, the
amount of entanglement is low and the information can be encoded in
tensors with small bond dimensions.

It may seem from the above argument that in the opposite limit of a
large (extensive) number of solutions, the bond dimensions would
necessarily be large. However, this is not true in general. Consider
the open boundary condition under which every locally compatible
configuration is a solution. In this extreme limit, the quantum
state~(\ref{eqn:quantum}) is an equal amplitude superposition of all
configurations, i.e., a product state. Such a state can be represented
with tensors of bond dimension one in the `$x$-basis'. One thus
expects that in cases of many solutions, the state should also be
close to a product state with low entanglement, and hence can be
represented with tensors of small bond dimensions. The above arguments
indicate that, if there is a highly entangled regime where the bond
dimensions required to represent the solution are large, then it must
necessarily be for systems with an intermediate number of
solutions. In Sec.~\ref{sec:results}, we show numerically that, even
in the intermediate regime where the solutions of an arbitrary vertex
model are more than just a few, it is possible to obtain an efficient
and compressed tensor network representation of the
allowed-configuration manifold.

Having argued that, for the case of problems with a small number of solutions, solutions of vertex models with partially fixed
boundaries can be encoded into tensor networks with small bond
dimensions, we set out to find this tensor-network representation. The
pertinent motivating question is: given that there is a representation
that can compress the full information of all solutions with
relatively small bond dimensions, how can we find it efficiently?

\section{Application of the ICD to the Random Vertex Model}
\label{sec:random}

In this section we apply ICD to
$N$ random instances of vertex lattices of fixed concentration of 
TOFFOLI gates, and fixed and equal concentrations of all four other types of gates
shown in Fig.~\ref{fig:vertex_gate}, with random input states for each instance.
By evaluating the full tensor trace for each of these $N$ instances and for various lattice sizes, we obtain information about the {\it average} scaling of performing the underlying classical computations by means of the ICD method. Moreover, we study the full distribution of the maximum bond dimension $\chi$ over random realizations and find that the {\it typical} behavior is generally different than the average, due to the presence of heavy tails in the bond-dimension distribution. Finally, we establish numerically that the scaling of the actual running time $\tau$ with the maximum bond dimension is always better than the worst-case estimate $\tau \sim \mathcal{O}(\chi^7)$.

\subsection{Local moves}
\label{sec:local_moves}

Since the vertex model is defined on a tilted square lattice, we first
need to turn it into a lattice as shown in Fig.~\ref{fig:tensor_net}
in order to apply our algorithm in Sec.~\ref{sec:algorithm}. This can
be done by performing local moves on the tilted lattice, which we
explain below.

\begin{figure}[hbt]
\centering
\includegraphics[width=.47\textwidth]{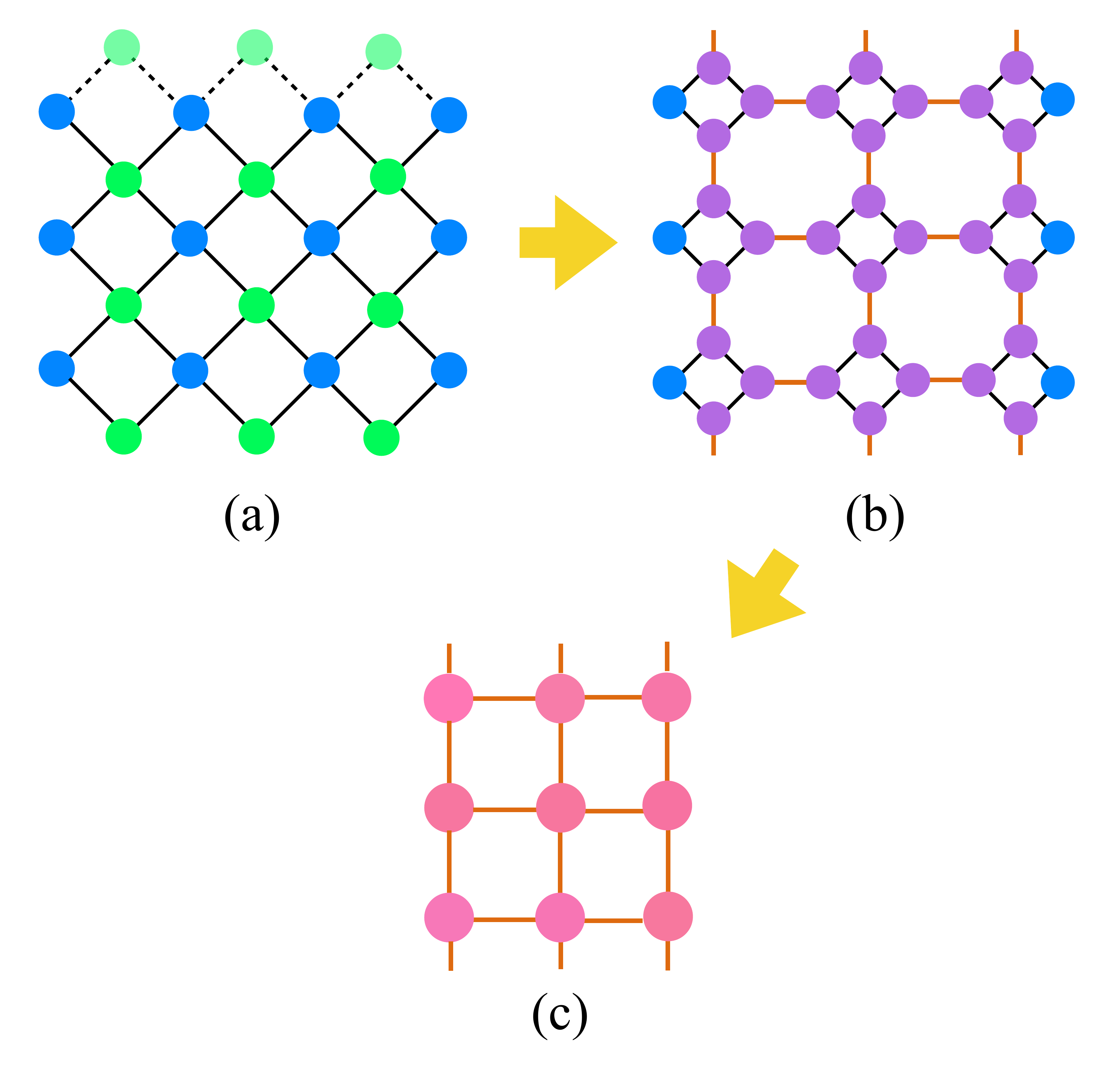}
\caption{(Color online) Illustration of the local moves which turn the
  original lattice into a square lattice rotated by $45^{\circ}$. In
  (a), sites belonging to sublattice $A$ and $B$ are shown in blue and
  green dots, respectively. From (b) to (c), four sites belonging to a
  diamond are contracted into one.}
\label{fig:move_lattice} 
\end{figure}

\begin{figure}[hbt]
\centering
\includegraphics[width=.45\textwidth]{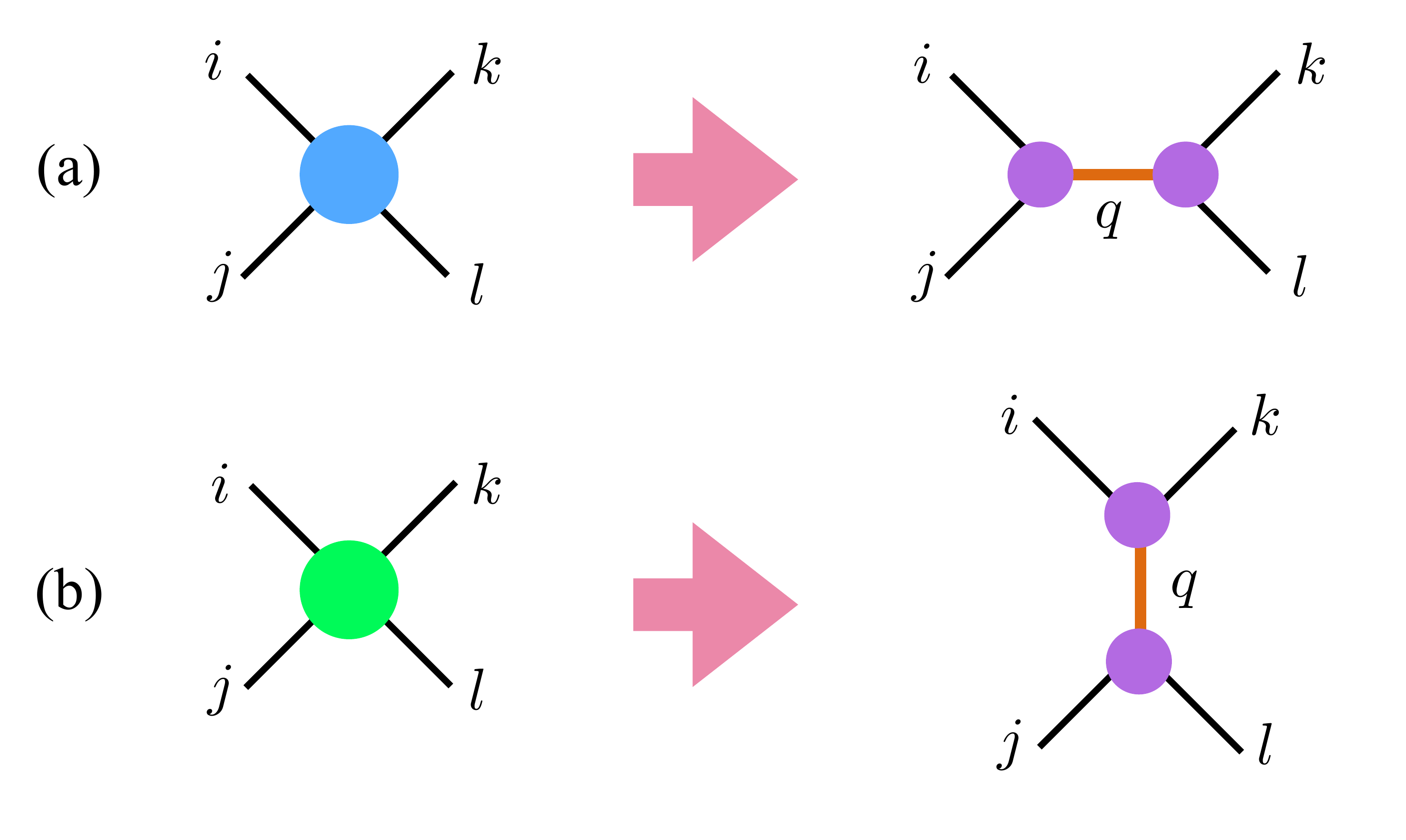}
\caption{(Color online) Local moves that decompose each tensor on the
  original lattice into two along either horizontal or vertical
  direction, depending on whether the site belongs to sublattice A (a)
  or B (b).}
\label{fig:local_move} 
\end{figure}

The tilted square lattice Fig.~\ref{fig:move_lattice}a is bipartite,
with two sublattices A and B. Local tensor decompositions and
contractions for tensors on each sublattice can rearrange the lattice
into an ``untilted'' one, rotated by $45^{\circ}$ with respect to the
original lattice. We start by splitting each tensor on the original vertex
lattice into two along either horizontal or vertical direction,
depending on which sublattice the corresponding site belongs to. Let
us take a bulk tensor $T_{ijkl}$ on the original lattice. If the site
belongs to sublattice $A$, we decompose the tensor horizontally into
two rank-3 tensors, $T_{ijkl}=\sum_q A_{ijq}\, B_{klq}$; if the site
belongs to sublattice $B$, we instead decompose the tensor vertically,
$T_{ijkl}=\sum_q \widetilde{A}_{ikq}\, \widetilde{B}_{jlq}$, as shown
in Fig.~\ref{fig:local_move}. Such a decomposition can be achieved via
an SVD on the original tensors, $T_{(ij), (kl)} = U_{(ij), q}\,
\Lambda_{q}\, V^\intercal_{q,(kl)}$ to yield $A_{ijq} = U_{(ij),q}\,
(\Lambda_{q})^{1/2}$ and $B_{klq} = (\Lambda_{q})^{1/2}\,
V^\intercal_{q,(kl)}$. We visit each site and split the tensors in
this way. This turns the tensor network into the structure shown in
Fig.~\ref{fig:move_lattice}b. We then further contract four tensors in
a diamond into one and finally arrive at a new square lattice rotated
by $45^{\circ}$ with respect to the original one
(Fig.~\ref{fig:move_lattice}c). With these local moves, which have to
be carried out only once, we cast the problem into the form discussed
in Sec.~\ref{sec:vertex}.

However, instead of doing an SVD on the
original tensor, here we can use the fact
that the tensors encode the truth tables of reversible gates and use
an alternative method. Define a new set of tensors with an auxiliary
index $q=0, 1, \dotso, 7$ labeling the vertex state,
$\widetilde{T}^q_{ijkl}$. Now the component of this rank-(4,1) tensor
is one if and only if $q$ is the same as the input state labeled by
$(i,j)$. Then, the desired decomposition can be achieved as follows:
\begin{eqnarray}
A_{ijq}=\sum_{kl}\widetilde{T}^q_{ijkl},
\ \ \ B_{klq}=\sum_{ij}\widetilde{T}^q_{ijkl}, \nonumber
\\ \widetilde{A}_{ikq}=\sum_{jl} \widetilde{T}^q_{ijkl},
\ \ \ \widetilde{B}_{jlq}=\sum_{ik}\widetilde{T}^q_{ijkl}.
\end{eqnarray}
One can easily check that the contraction of the $A$ and $B$ tensors
gives back the original tensor $T$, and hence this achieves the
splitting shown in Fig.~\ref{fig:local_move}. The remaining steps of
the algorithm are carried out exactly in the same way as before. By
construction, the bonds between the resulting bulk tensors all have
dimension 8.

\subsection{Control of bond dimensions}

We can now apply the compression-decimation algorithm to count the
number of solutions for a given boundary condition. As we discuss in
Sec.~\ref{sec:algorithm-sweeps}, a truncation threshold $\delta$ needs
to be specified in the sweeping step of the algorithm. Since we are
performing an \textit{exact} counting, no approximation in the
truncation of the bond dimensions is made during the coarse-graining
procedure, i.e., we choose $\delta=0$ within machine precision. This
is a key methodological difference of the ICD to TNRG methods, which
\textit{approximate} physical observables to within a certain accuracy
by enforcing a finite $\delta$.

As mentioned above, from a statistical mechanics point-of-view, what
we are computing is the zero-temperature partition function of the
vertex model, which yields the ground state degeneracy. In the bulk,
all locally compatible configurations are equally possible until they
receive information from the boundary conditions. Therefore, the
coarse-graining step effectively brings the boundaries close to one
another, and the sweeping step propagates information from the
boundary to the bulk and knocks out states encoded in local tensors
that are incompatible with the global boundary conditions.

The reason why the growth of bond dimensions remains controlled is
that longer-range compatibility constraints over increasingly larger
areas are enforced upon the coarse-grained tensors. These constraints
are propagated to neighboring coarse-grained tensors upon sweeping,
thus further reducing bond dimensions and compressing the
tensor-network representation. For the trivial cases of either fixing
all gates on one boundary or leaving them all free (open boundary
condition), we have checked that the tensors converge to bond
dimension one (scalars) after one sweep, \textit{without the need of
  coarse-graining}. The tensor contraction is then simply reduced to
multiplications of scalars, which can be trivially computed and indeed
gives the correct counting. This demonstrates that the sweeping is
responsible for propagating information from the boundary, and that
the case of fully fixing one boundary is thus equivalent to direct
computation, as described in Sec.~\ref{sec:toffoli}.

In cases of mixed boundary conditions, the sweeping on the original
lattice scale will generally not be sufficient to propagate
information across the whole system or establish full communications
between the two boundaries. Thus one would expect that while the bond
dimensions close to the boundary may be small, those deep in the bulk
may be large. We therefore perform the contractions selectively on
rows and columns containing mostly tensors with small bond dimensions
while leaving the rest for the next coarse-graining step, as described
in Sec.~\ref{sec:algorithm}.

\subsection{Numerical results}
\label{sec:results}

The computational cost of the ICD algorithm is determined by the
maximum bond dimension encountered during the coarse-graining and
sweeping procedures. In this section, we study the scaling of the
maximum bond dimension as function of the set of parameters defining
an instance of the problem: the number of vertices in each column $L$,
the total number of columns (circuit depth) $W$, the concentration of
TOFFOLI gates $c$, and the number of unfixed boundary vertices
$L_{\partial}$. For a given set of parameters, we consider random
tensor networks corresponding to typical instances of computational problems. 
By looking at the scaling of the bond dimensions, we gain some
understanding of how the hardness of the problems depends on various
parameters, which may serve as a guidance for designing and analyzing
computational circuits for practical problems. 

\begin{figure}[t]
\centering
\includegraphics[width=.49\textwidth]{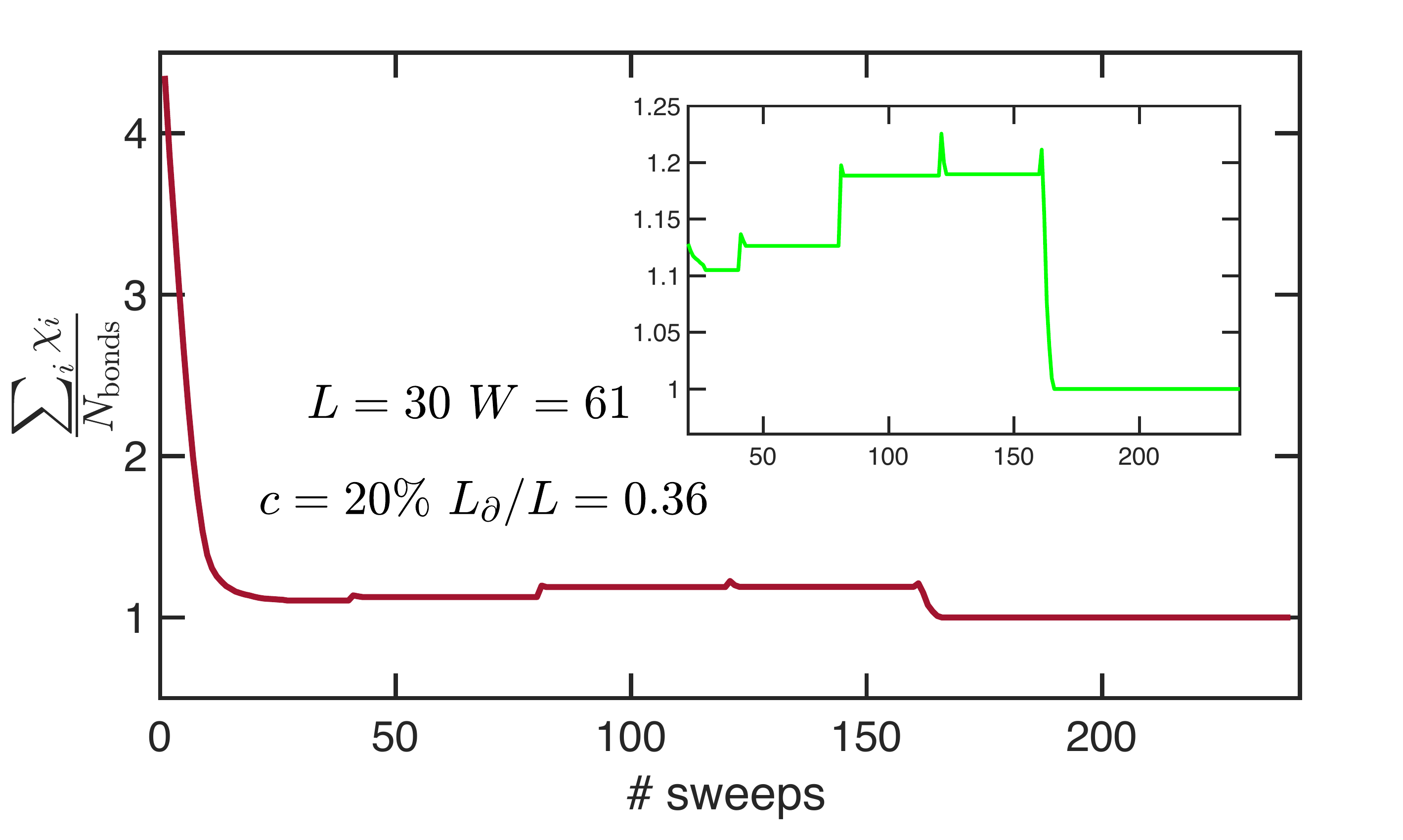}
\caption{(Color online) The average bond dimension of the entire
  lattice as a function of the number of sweeps in the
  compression-decimation steps. The bumps where the average bond
  dimension increases slightly correspond to the points where we
  coarse-grain the lattice via column and row contractions. Inset:
  zoom-in plot from the 20th sweeping step.}
\label{fig:bond_dim_sweep} 
\end{figure}

Before looking into the scaling of the maximum bond dimensions, we
first show the average bond dimension for the entire lattice as a
function of the number of sweeps in the compression-decimation
steps. As seen from Fig.~\ref{fig:bond_dim_sweep}, the average bond
dimension indeed decreases as the sweeping is performed. The bumps in
the plot correspond to the points where we coarse-grain the lattice
via column and row contractions. At a given length scale, the average
bond dimension converges after a few sweeps. As we increase the length
scales, the average bond dimension may first increase, but will
eventually drop again as we perform sweeps at the new length
scale. This demonstrates that the sweeping is able to impose global
constraints at the boundary into the bulk, hence keeping the bond
dimensions of bulk tensors under control.

\begin{figure}[t]
\centering
\includegraphics[width=.47\textwidth]{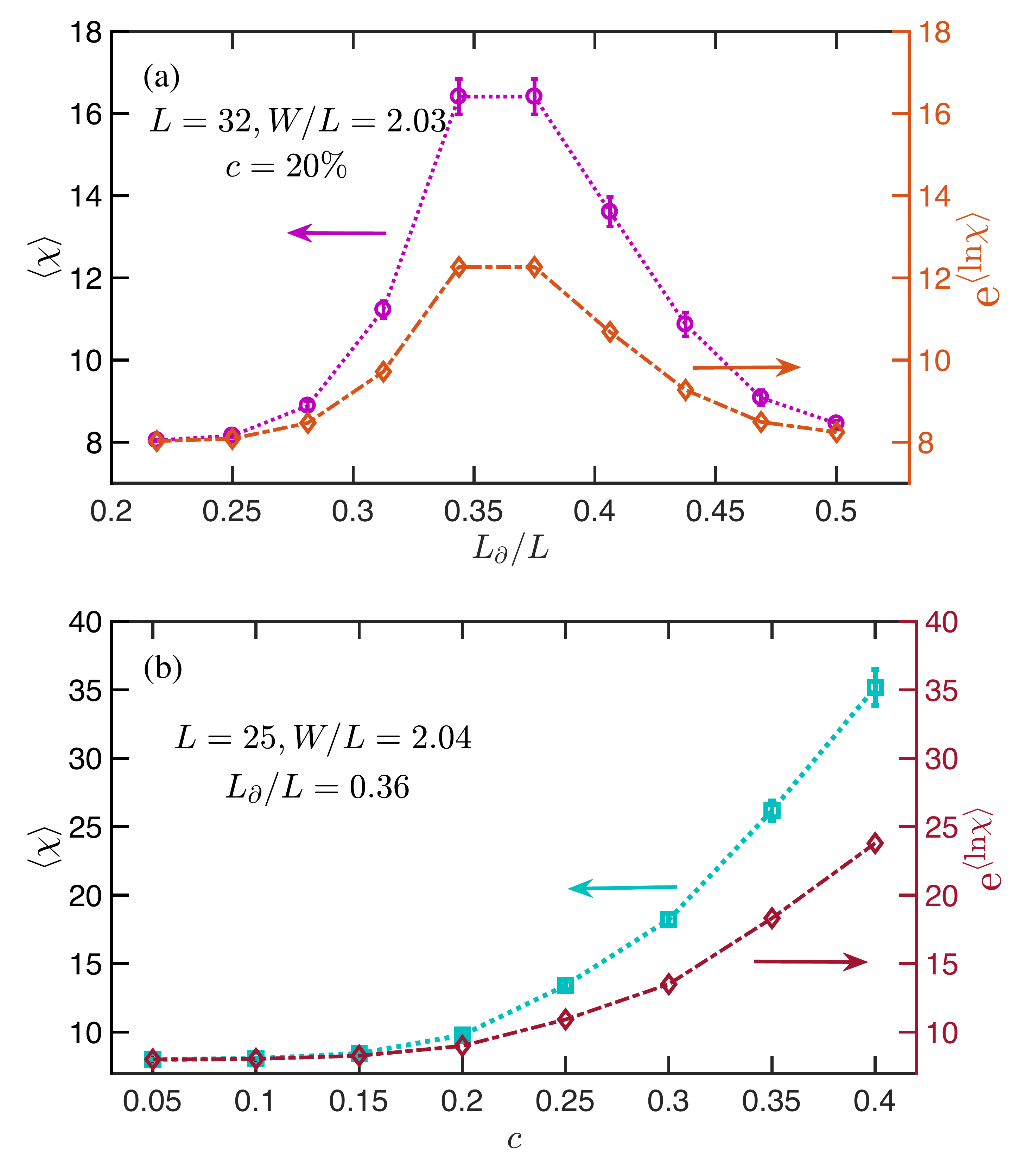}
\caption{(Color online) Scaling of the average maximum bond dimension $\chi$
  with (a) the ratio of unfixed boundary vertices $L_{\partial}/L$,
  and (b) TOFFOLI concentration $c$, versus the scaling of the typical maximum bond dimension $e^{\langle {\rm ln}\chi \rangle}$. The remaining parameters are
  fixed in each plot. The data are obtained by averaging over 2000
  realizations of random tensor networks.}
\label{fig:scaling_1} 
\end{figure}

We expect the maximum bond dimension to follow the scaling function
$\chi = \mathcal{G}(L_{\partial}/L, c, L, W/L)$. Below we study the
growth of maximum bond dimensions as a function of each system
parameter numerically. First, we consider the scaling of $\chi$ as the
ratio of unfixed boundary vertices $L_{\partial}/L$ is varied, with
the other parameters fixed. As shown in Fig.~\ref{fig:scaling_1}a, the
bond dimensions are small for both small and large
$L_{\partial}/L$. This is in agreement with our discussions in
Sec.~\ref{sec:entanglement}, where we argued that in both regimes the
states are close to product states and there should exist a
representation in which the bond dimensions are small (the `$z$-basis'
and `$x$-basis'). For intermediate values of $L_{\partial}/L$, the
bond dimensions grow, indicating the existence of a hard regime where
either there is no such a representation of small bond dimensions to
fully encode the solutions, or it is very hard to find such a
representation via tensor optimization algorithms.

\begin{figure*}
\centering
\includegraphics[width=0.9\textwidth]{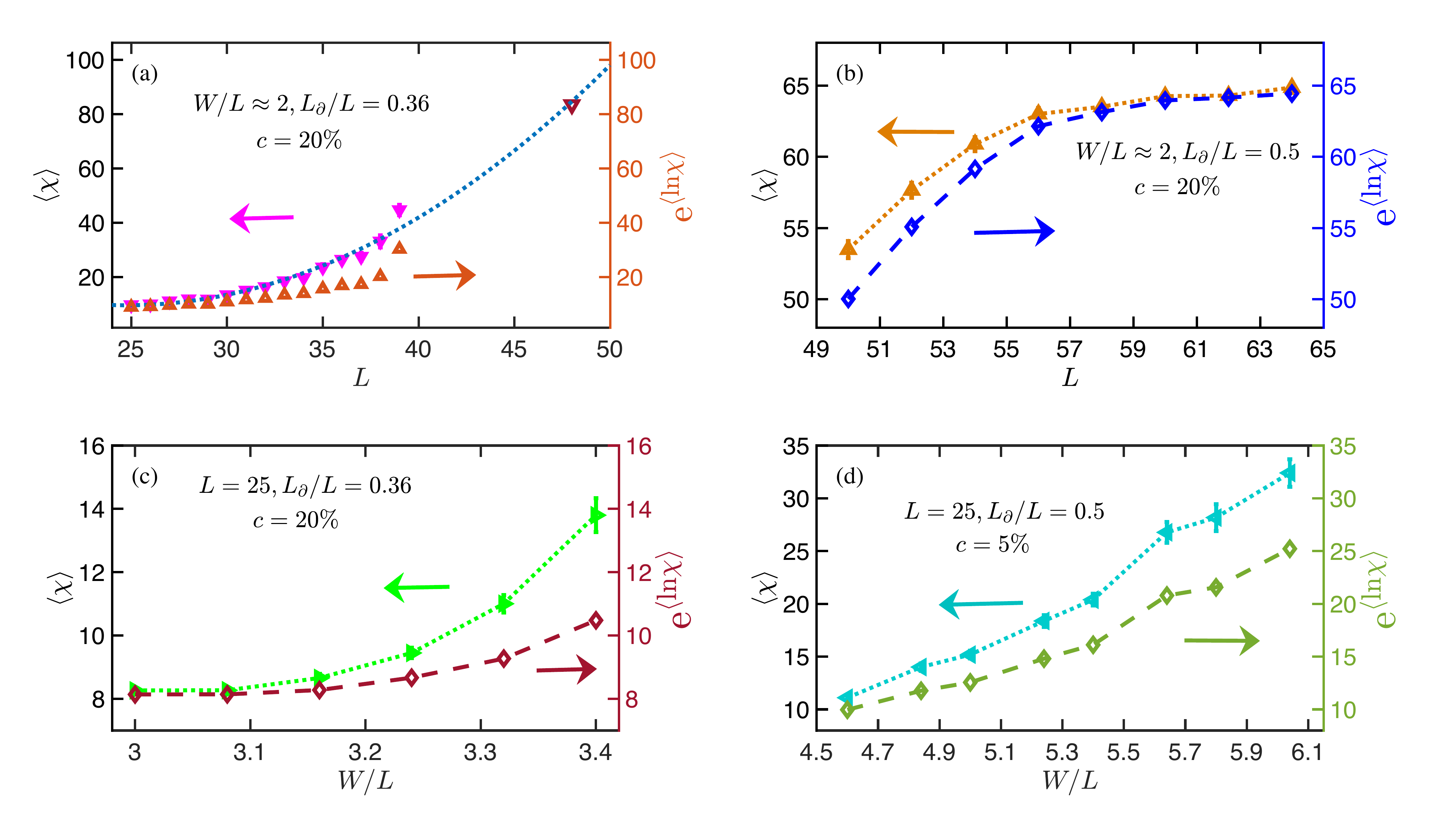}
\caption{(Color online) Scaling of the average maximum bond dimension $\langle \chi \rangle$
  with $L$ (a,b) and $W/L$ (c,d), versus the scaling of the typical maximum bond dimension $e^{\langle {\rm ln}\chi \rangle}$. The data are obtained by averaging
  over 500 to 2000 realizations of random tensor networks. In (a), the
  last point is averaged over 7 realizations and the error bar is not
  shown. The blue dotted line is a guide to the eye, and corresponds
  to a quadratic fitting. In all cases, the typical values stay below the average values, due to the presence of heavy tails in the distribution of $\chi$.}
\label{fig:scaling_2} 
\end{figure*}

Fixing $L_{\partial}/L=0.36$, which corresponds to the hard regime in
Fig.~\ref{fig:scaling_1}a, we plot the scaling of $\chi$ as a function
of the TOFFOLI concentration $c$. The TOFFOLI gates impose nontrivial
vertex constraints, which involve a nonlinear relationship between the
input and output bits. In fact, in the absence of TOFFOLI gates, the
vertex model can be expressed as $3L$ decoupled Ising chains whose
dynamics are simple~\cite{chamon}. In the ICD algorithm,
the maximum bond dimension indeed grows with increasing TOFFOLI
concentrations, as depicted in Fig.~\ref{fig:scaling_1}b.

Now let us look at the scaling of $\chi$ as a function of the input
size $L$. Again, we fix $L_{\partial}/L=0.36$ to stay in the hard
regime. Figure~\ref{fig:scaling_2}a shows that the maximum bond
dimension increases with increasing input size, even when the aspect
ratio $W/L$ of the circuit is fixed. Because of the limited range of
$L$ we were able to analyze, we cannot draw any conclusion regarding
the functional form of this scaling, which would determine the
complexity of our algorithm. However, we can demonstrate that our
algorithm is able to solve the problem in regimes that are still
intractable using a naive enumeration of solutions. For this purpose,
we move away from the hardest regime and choose
$L_{\partial}/L=0.5$. As can be seen in Fig.~\ref{fig:scaling_2}b, we
are able to reach much larger values of $L$ in this regime, and the
bond dimensions, although still growing, increase at a much slower
pace. In fact, we were able to reach $L=96$ with an average maximum
bond dimension $\langle \chi \rangle = 78.25$ (data not shown). Since
half of the input vertices are unknown, a direct trial-and-error
enumeration would take $8^{48} \approx 10^{43}$ iterations to perform
an exact counting, which is prohibitive even with parallelization. We
have thus shown that there is a subset of nontrivial problems that can
easily be solved by the ICD method, for which (a) direct enumeration
is impossible to scale up and (b) efficient custom algorithms are not
known.

Finally, we show the scaling with the aspect ratio $W/L$. As we
previously discussed, the key to the reduction of the bond dimensions
of bulk tensors is the global constraint imposed at the boundary. The
coarse-graining step brings the boundaries close together while the
sweeping step helps propagate information. Therefore, one should
expect the problem to become harder as the circuit depth $W$ increases
for a fixed $L$, since it takes more iterations of coarse-graining for
the connection between the boundaries to be built, all the while the
bond dimensions of the bulk tensors barely decrease. In
Fig.~\ref{fig:scaling_2}c we show the scaling of $\chi$ as a function
of $W/L$ in the hard regime; $\chi$ indeed grows upon increasing
$W/L$, as expected. For computational problems of practical interests,
the vertex model representation often has the feature of low TOFFOLI
concentration but large aspect ratio, e.g., the multiplication
circuit~\cite{chamon, vedral}. In Fig.~\ref{fig:scaling_2}d we show
data for cases with this feature by lowering the TOFFOLI concentration
to $c=5\%$ and keeping $L_{\partial}/L=0.5$. We find that the bond
dimensions grow in a similar fashion as in Fig.~\ref{fig:scaling_2}c,
although a larger range of values of $W/L$ now becomes amenable.

\begin{figure}[t]
\centering
\includegraphics[width=.5\textwidth]{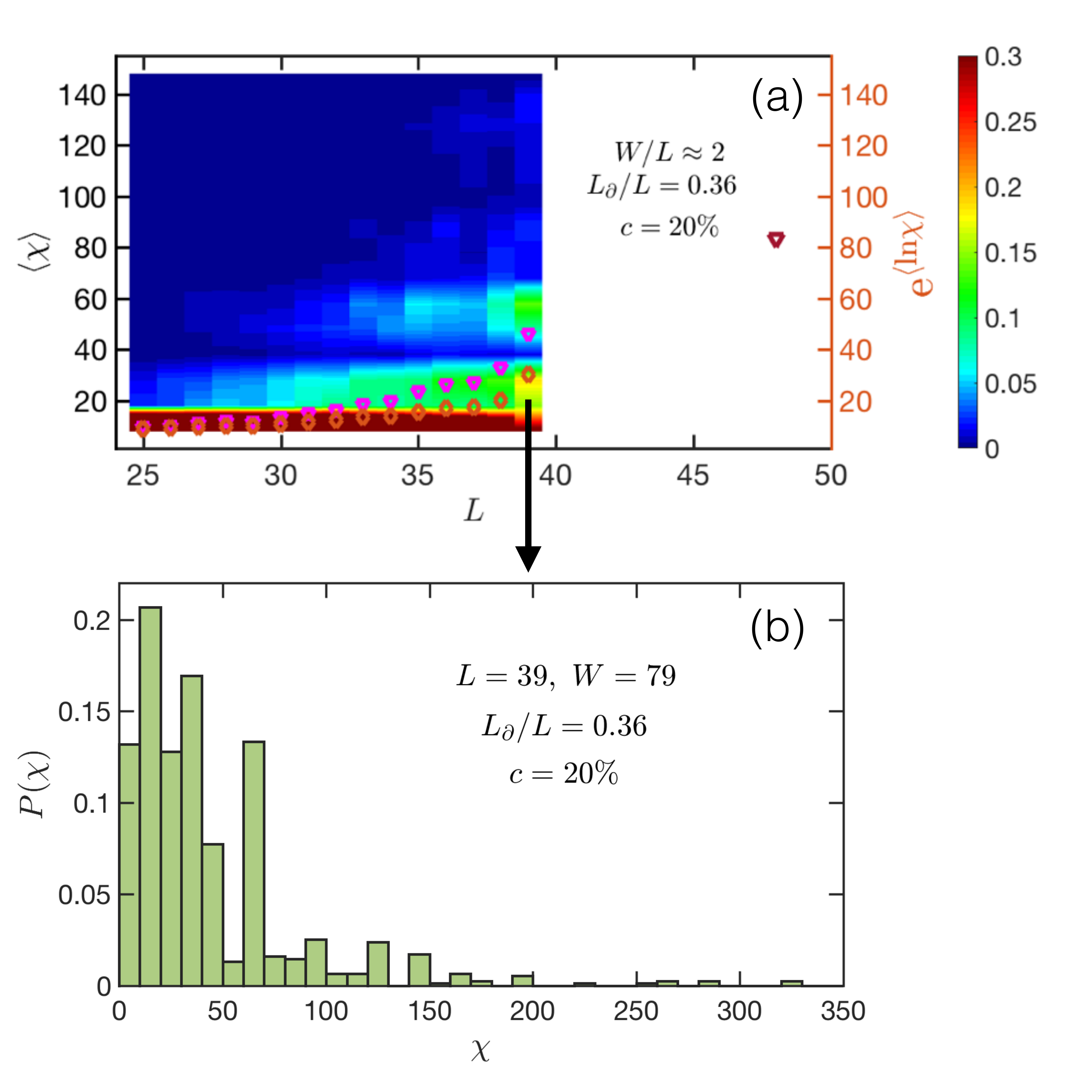}
\caption{(Color online) The full distribution of the maximum bond dimension over random instances. In (a), the color plot along the vertical direction shows the probability distribution at each $L$. The orange (diamond) points show the values of $e^{\langle {\rm ln} \chi \rangle}$, giving an estimate of the typical instances in contrast to the average instances depicted in purple (triangle) points. In (b), we take the slice of $L=39$ in (a) and plot the histogram of the distribution.}
\label{fig:distribution} 
\end{figure}

The above results demonstrate the {\it average} scaling behavior of the ICD algorithm over random instances of computations. It is informative to compare this to the {\it typical} behavior, revealed by analyzing the full distribution of the maximum bond dimensions; see, e.g., Fig.~\ref{fig:scaling_2}(a). In Fig.~\ref{fig:distribution}, we present the probability distribution of the maximum bond dimension. A vertical cut at each $L$ corresponds to the probability distribution over all instances for that $L$. For larger $L$ we observe secondary peaks at larger $\chi$, which gradually take up more weight, thus shifting both average and typical maximum bond dimension to higher values. Moreover, despite the fact that the highest weight is always encountered at small bond dimensions, a finite subset of hard instances generate much larger bond dimensions, leading to heavy tails in the distributions. Average values are sensitive to such tails, and hence do not faithfully represent the typical instances. In Fig.~\ref{fig:scaling_1} and \ref{fig:scaling_2}, we also plot the values of $e^{\langle {\rm ln} \chi \rangle}$ as an estimate of the typical behavior, in contrast to $\langle \chi \rangle$. Indeed, the typical values stay below the average values in all cases studied. We point out that the presence of heavy tails in the distribution is ubiquitous in random satisfiability problems, and such instances could in principle be tackled with different strategies~\cite{crosson, troyer, wecker, yang}.

\begin{figure}[t]
\centering
\includegraphics[width=.47\textwidth]{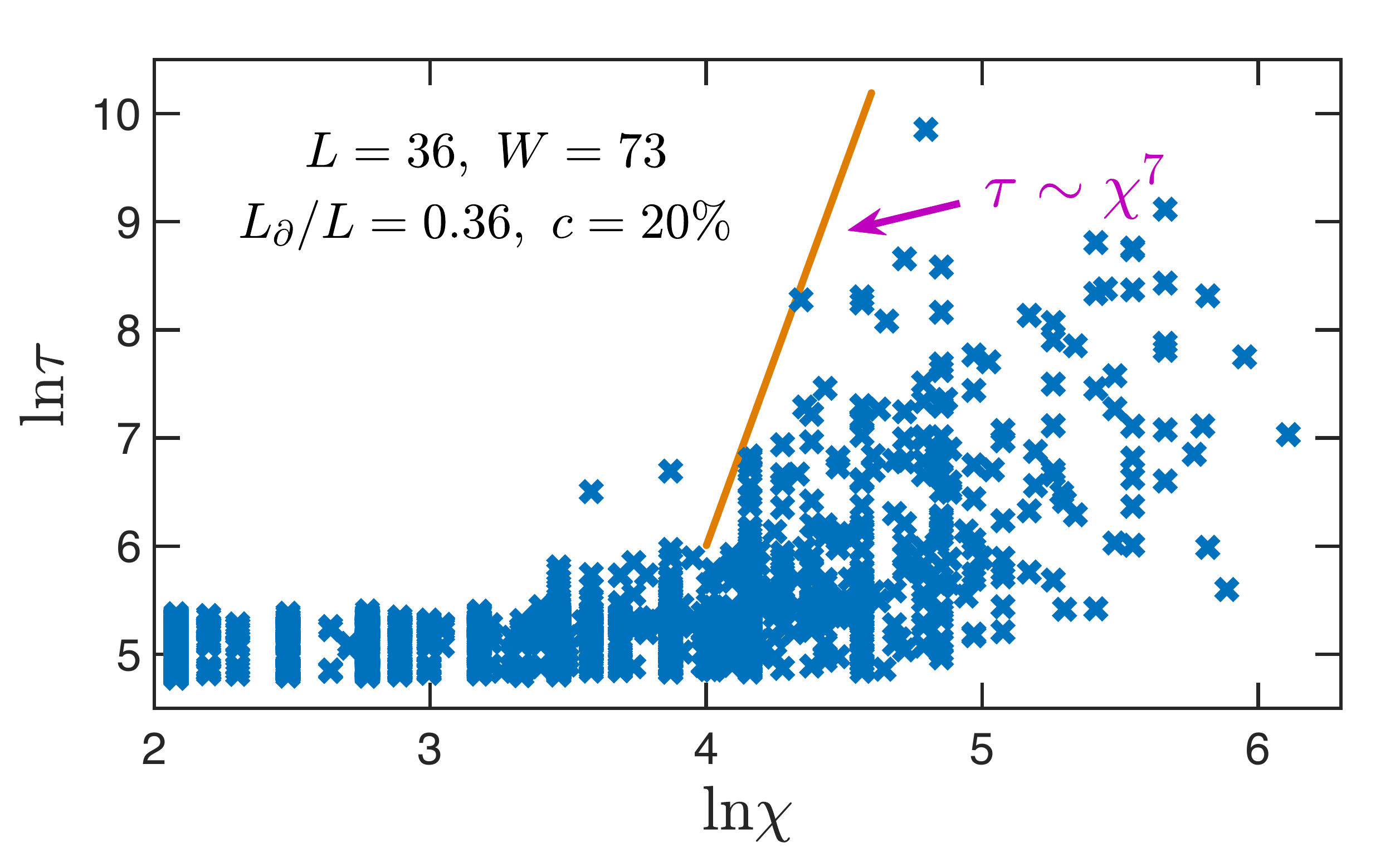}
\caption{(Color online) Scatter plot of the (logarithm of) actual running time $\tau$ in units of seconds versus the (logarithm of) maximum bond dimension $\chi$ for 4600 instances. The calculations were performed using a Python implementation of the ICD algorithm, using NumPy / LAPACK for all linear algebra operations, on 2.0 GHz Intel Xeon Processors E7-4809 v3.}
\label{fig:calibration} 
\end{figure}

The efficiency of the ICD algorithm is controlled by the maximum bond dimension encountered in each instance, and in particular, the complexity of the algorithm is upper bounded by $\mathcal{O}(\chi^7)$ as discussed in Sec.~\ref{sec:algorithm}. Nevertheless, it is still useful to see whether the actual running time saturates this bound. In Fig.~\ref{fig:calibration} we show the scatter plot for the actual running time $\tau$ versus $\chi$ for 4600 random instances. We see a clear clustering of the data points and a positive correlation between these two quantities. The fact that there is a spreading of $\tau$ for each $\chi$ can be understood by taking into account the nonuniform spatial distributions of the bond dimensions across the system. Unlike the TNRG algorithms, where
the bond dimensions of all tensors and all tensor legs are frequently chosen to be uniform, bond dimensions of different tensors and of different
legs of the same tensor are typically highly nonuniform in the ICD method. Therefore, running times for instances with the same maximum $\chi$ also depend on the number of bonds with dimension $\chi$. The distribution of $\chi$
throughout the system is thus an important factor. Moreover, we find that the scaling of the running time with the maximum bond dimension $\tau \sim \chi^{\alpha}$ has a power $\alpha < 7$, which shows that the actual performance of the algorithm is generally better than the worst-case scenario estimate.

\section{Summary and outlook}
\label{sec:outlook}

We presented a method for contracting tensor networks that is well
suited for the solution of statistical physics vertex models of
universal classical computation. In these models, the tensor trace
represents the number of solutions. Individual solutions can be
efficiently extracted from the tensor network when the number of
solutions is small.  More generally, the method applies to any system,
classical or quantum, whose quantity of interest is a tensor trace in
an arbitrary lattice.

Our scheme consists of iteratively compressing tensors through a
contraction-decomposition operation that reduces their bond
dimensions, followed by decimation, which increases bond dimensions
but reduces the network size. By repeated applications of this two
step process -- compression followed decimation -- one can gradually
collapse rather large tensor networks.

In the context of computation, the method allowed us to study
relatively large classical reversible circuits represented by two
dimensional vertex models. By contrast with thermal annealing, direct
computation from a fully specified input boundary through the use of
tensor networks occurs in a time linear in the depth of the
circuit. For complex problems with partially fixed input/output
boundaries tensor networks enable us to count solutions in problems
where enumeration would otherwise take of order $8^{50}$ operations.

We close with an outlook of future directions motivated by this work.

First, focusing on the method {\it per se}, the performance of our ICD
algorithm could still be further improved. There are enhancements that
are simply operational in nature, such as parallelization of the
sweeping step of the algorithm, which can be accomplished by dividing
the tensors into separate non-overlapping sets.

Second, at a more fundamental level, as we point out at the end of
Sec.~\ref{sec:results}, a better understanding of the mechanism by
which short-range entanglement is removed within the ICD method would
require a systematic study of the evolution of the spatial
distribution of bond dimensions. The goal would be to design more
controlled bond dimension truncation schemes that involve the effect
of the environment of local tensors, as proposed in Refs.~\cite{wen,
  wen2, xiang1, xiang2, Evenbly2017}. More generally, we expect that
our method can be applied to both classical and quantum many-body
systems in two and higher dimensions.

Third, in our study of computation-motivated problems, we focused on
random tensor networks corresponding to random computational
circuits. However, the ICD methodology should be used to address
problems of practical interest, a research direction that is being
currently explored. The results on the scaling of the bond dimensions
presented above should inform the design and analysis of tractable
computational circuits, such as circuits with $W\sim L$ and a moderate
number of TOFFOLI gates. Multiplication circuits based on partial
sums, for instance, are very dense in TOFFOLI gates, and hence are not
good {\it a priori} candidates for tensor network formulations of
related problems, such as factoring. However, different multiplication
algorithms whose associated vertex models are less dense in TOFFOLI
gates, and other computational problems could be amenable by our
approach.  Identifying classes of computational problems of practical
interest that can be tackled with tensor network methods remains an
open problem at the interface between physics and computer science.

Finally, from a statistical mechanics point-of-view, one may speculate
that the ICD algorithm could allow us to study the glass phase of
disordered spin systems for which classical Monte Carlo dynamics
breaks down due to loss of ergodicity.
 
\section*{Acknowledgments}
We thank Justin Reyes, Oskar Pfeffer, and Lei Zhang for many useful
discussions. The computations were carried out at Boston University's
Shared Computing Cluster. We acknowledge the Condensed Matter Theory
Visitors Program at Boston University for support. Z.-C. Y. and
C. C. are supported by DOE Grant No. DE-FG02-06ER46316. E. R. M. is
supported by NSF Grant No. CCF-1525943.

\bibliography{manuscript}

\end{document}